# Analysis Of Critical Power Loss In A Superconductor

**Mario Rabinowitz**

*Stanford Linear Accelerator Center, Stanford University, Stanford, California 94305*

Inquiries to: *Armor Research, 715 Lakemead Way, Redwood City, CA 94062*
Mario715@earthlink.net

**Abstract**

A critical power dissipation resulting from an oscillating magnetic field, $H_p \cos \omega t$, can produce a magnetic breakdown field, $H_p^{'} < H_c$, the critical field of the superconductor. The analysis shows, for example, why the breakdown field of a superconducting microwave cavity can be well below $H_c$ in some cases, and indicates what the functional dependence of the cavity Q may be for values of $H_p$ near $H_p^{'}$. The effective resistivity of a single isolated oscillating fluxoid, as well as that of a stationary normal region, is also derived for both type I and type II superconductors.

## I. INTRODUCTION

It has been observed in both low and high frequency measurements that the onset of excessive power dissipation in superconductors can occur at values of the peak applied magnetic field $H_p^{'}$, below $H_c$, the critical magnetic field ($H_{c1}$ for type II). This has been variously ascribed to local magnetic field enhancement due to surface roughness, presence of impurities, etc. Easson et al. [1] suggested "that the superconducting to normal transition in type II superconductors is caused by a temperature rise above $T_c$ due to ac losses, rather than by the peak ac currents in the sample rising to the thermodynamic critical value." Halbritter [2] recognized that this suggestion for the low frequency case might be applicable to superconducting microwave cavities. As far as could be determined from the available literature, it appears that neither Halbritter, Easson et al., nor anyone else, has pursued this suggestion in terms of a theoretical analysis which relates the magnetic breakdown field to the thermal and electrical properties of a superconductor.

The main object of this paper is to calculate the magnetic breakdown field, in terms of a power-dissipation and thermal-conduction analysis. Two cases will be considered: that of a normal region parallel to, and that of a normal region perpendicular to, the surface of a superconductor.

**A. Normal Region Parallel to Surface**

For generality, consider a superconductor of thickness f on a normal substrate of thickness t, where $f > \lambda$, the penetration depth. The results will then also be applicable to a bulk superconductor with $t = 0$ and $f =$ bulk thickness. Assume that a normal region of radius $a$, such as a fluxoid, lies parallel to, and its axis is a distance d from, the surface of the superconductor, and a distance b from the outer surface of the substrate, as shown in Fig. 1.

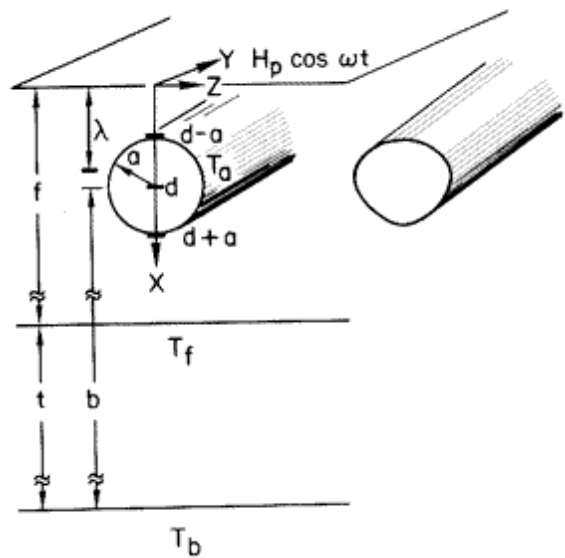

Fig.1.Schematic representation of two fluxoids parallel to the surface. The fluxoid on the right indicates the "pear-shaped" cross section resulting from a temperature gradient across it during power dissipation.

Let us now consider an oscillating fluxoid which, as we shall see in See. IV-A, is equivalent to a stationary normal region with the supercurrent, $j_z$, through it, and which gives the largest power loss. Due to the magnetic field $H_p \cos\omega t = \text{Re}[H_p \exp(i\omega t)]$, applied at the surface of the superconductor, the average power dissipation per unit volume in the normal region is

$$dP'/dV = \rho \langle j_z^2 \rangle, \tag{1.1}$$

where p is the effective resistivity of the normal region (to be derived in Sec. IV) and the current density is

$$\begin{aligned}j_z &= j_o \exp(-x/\lambda)\exp(i\omega t) \\ &= (H_p/\lambda)\exp(-x/\lambda)\exp(i\omega t)\end{aligned} \tag{1.2}$$

Therefore, the average power loss per unit length is:

$$\begin{aligned}P &= \tfrac{1}{2}\rho j_o^2 \int_{d-a}^{d+a} 2\exp(-2x/\lambda)\left[a^2-(x-d)^2\right]^{1/2} dx \\ &= \tfrac{1}{2}\rho(H_p/\lambda)^2 \left(0.886\pi^{1/2} a\lambda\right)\exp(-2d/\lambda) I_1(2d/\lambda)\end{aligned} \tag{1.3),(1.4}$$

where $I_1$ is the modified Bessel function of order one. The average power loss per unit area around the normal region is:

$$P_a = /2\pi a = \tfrac{1}{2}(\rho/2\pi\lambda)H_p^2[1.57\exp(-2d/\lambda)I_1(2a/\lambda)] = \tfrac{1}{2}R(FH_p)^2, \tag{1.5}$$

where

$$R = \rho/2\pi\lambda \tag{1.6}$$

is the effective normal state surface resistance of the cylinder. $R = R_o + R(T) \approx R_o$, as the temperature excursion is not great; and

$$F = \exp(-d/\lambda)[1.57 I_1(2a/\lambda)]^{1/2}. \tag{1.7}$$

Even prior to magnetic breakdown, a fluxoid will grow in cross section and change its shape as shown in Fig. 1 as its temperature is increased. These effects will be neglected at this time.

For a stationary normal region with $a \leq \delta \equiv$ the anomalous skin depth, the normal current density, $j_n$, is essentially uniform throughout it. If $a > \delta$, then the analysis would be similar to that given here. For all cases, there is an equivalent F and R so that the heat-conduction analysis to follow is quite general.

A rough representation for the superconducting thermal conductivity (see Fig. 2) is [3]:

$$\begin{aligned} K_s &\sim k_1(T - T_1) & T_a \geq T \geq gT_c \\ &\sim k_2 T^{-3/2} + c_2 & gT_c \geq T \geq hT_c \\ &\sim k_3 T^3 & hT_c \geq T \geq hT_f \end{aligned} \quad (1.8)$$

where $T_a$ is the temperature at the periphery of the fluxoid, $T_c$ is the critical temperature, g and h are fractions where $1 > g \geq h > 0$, and $T_f$ is the temperature at the interface, neglecting the thermal boundary resistance temperature drop. (If the normal region is a fluxoid, $K_s$ requires modification due to the high self-magnetic field. Another complication is that the effective thermal conductivity may be reduced very near a free surface, from ordinary bulk values, when the mean free path of the heat carrier is long compared with the distance from the surface. The extent of this effect is related to the nature of the heat carrier and whether it is directly energized, or indirectly through collisions. However, these modifications should not alter the essential results.) For the normal substrate, the conductivity is $K_n \approx kT$ for $T_f \geq T \geq T_b$, where $T_b$ is the temperature of the outer surface of the normal conductor = bath temperature, neglecting Kapitza resistance. (Temperature gradients in the bath are negligible for $T_b < T_\lambda$.

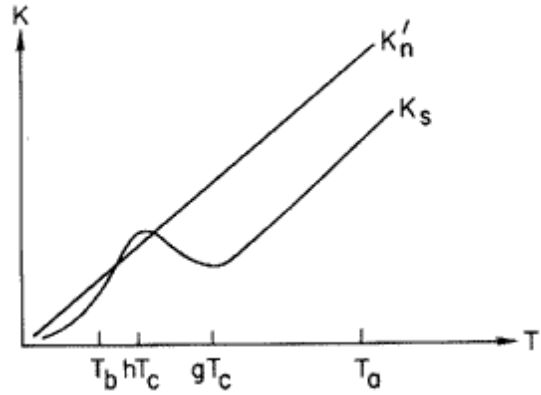

Fig. 2. Superconducting and normal thermal conductivity.

Assuming cylindrical symmetry as a fair approximation, and using cylindrical coordinates centered at the cylinder, the heat flow equation is $NP_a(a/r) = -K(dT/dr)$, where N is a factor to correct for the departure from cylindrical symmetry (see Fig. 3). Let b' = f - d. The solutions to the heat flow equation are:

$$\tfrac{1}{2} NRF^2 H_p^2 a \, \ell n(b'/a) = \tfrac{1}{2} k_1 \left(T_a^2 - g^2 T_c^2\right) + k_1 T_1 (gT_c - T_a) + 2k_2 \left[(hT_c)^{-1/2} - (gT_c)^{-1/2}\right] + c_2(gT_c - hT_c) + \tfrac{1}{4} k_3 \left[(hT_c)^4 - T_f^4\right] \quad (1.9)$$

and

$$\tfrac{1}{2} NRF^2 H_p^2 a \, \ell n(b'/b) = \tfrac{1}{2} k \left(T_f^2 - T_b^2\right). \quad (1.10)$$

It is assumed that the power dissipation in the rest of the superconductor results in a negligible temperature rise. If not, this incremental temperature can be added on. The term $k_1 T_1(gT_c - T_a)$ in Eq. (1.9) will be neglected to first approximation.

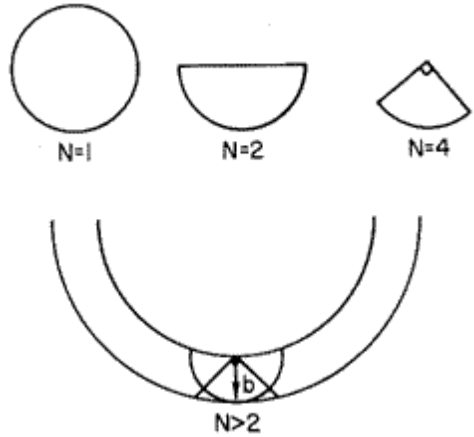

Fig.3. Schematic diagram to illustrate how the factor N corrects for the departure from cylindrical symmetry. The bottom drawing indicates a cross section of the cylindrical wall of a superconducting cavity, whose inside is evacuated and whose outside surface is in contact with the temperature bath.

Solving Eq. (1.9) for $H_p$, the result is:

$$H_p = F^{-1}\left[\frac{k_1\left(T_a^2 - g^2 T_c^2\right) + 2C_1'}{NRa\,\ell n(b'/a)}\right]^{1/2}, \qquad (1.11)$$

where

$$C_1' = 2k_2 T_c^{-1/2}\left(h^{-1/2} - g^{-1/2}\right) + c_2(gT_c - hT_c) + \tfrac{1}{4}k_3\left[(hT_c)^4 - T_f^4\right]$$

For $k_1 T_a^2 \gg k_1 T_m^2 - 2C_1'$,

$$H_p \approx F^{-1}[k_1/NRa\,\ell n(b'/a)]^{1/2} T_a\left\{-\tfrac{1}{2}\left[\left(k_1 g^2 T_c^2 - 2C_1'\right)/k_1 T_a^2\right] - \tfrac{1}{8}(...)^2 - ...\right\} \qquad (1.12)$$

A fluxoid parallel to the conducting wall is likely to be trapped near the inner conducting surface due to an incomplete Meissner-Ochsenfeld effect as the superconductor is cooled below its transition temperature. The reason for this is the low thermal conductivity and that cooling is generally initiated on the outside surface, which drives the fluxoid inward as the material undergoes transition. The thermodynamic potential due to the magnetic interaction between the fluxoid and the external field together with related screening currents may also drive the fluxoid toward the surface, or present a potential barrier near the surface depending on field level. Pinning centers (which may be enhanced near a surface) serve to keep the fluxoid(s) from being driven out, and to assume the same equilibrium position after each cool down. This may account for the reproducibility of $H_p'$, and possibly even of the residual Q.

## B. Normal Region Perpendicular to Surface

The analysis of a normal region perpendicular to the surface of a superconductor is more complicated for geometrical reasons as well as the fact that the heat is conducted to the temperature bath, $T_b$ through both normal and superconducting regions. We will consider only the case of a normal region (such as a fluxoid) of radius a, which is perpendicular to and goes completely through the wall of the superconductor of thickness b. As shown in Fig. 4, if the normal. region is a fluxoid, it will grow larger at the conducting surface due to the temperature gradient along its length during power dissipation. Growth prior to breakdown will be neglected at this time.

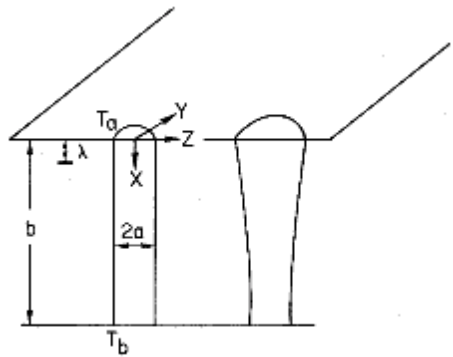

Fig.4. Schematic representation of two fluxoids perpendicular to the surface. The fluxoid on the right indicates the increase in cross section due to the temperature gradient during power dissipation.

For a pure, highly defect-free, material, the normal thermal conductivity $K_n' = k_{1n}T$ is $> K_s$ over most of the temperature range [4], as depicted in Fig. 2. In addition, the electrons are directly energized by the electromagnetic field and do not need to transfer power to the phonons in order to transport it to the heat sink. Therefore, to a rough approximation, let us assume that all the heat is conducted (channeled) down the normal cylinder. The actual temperature rise will be less than the calculated temperature $T_a$ at the conducting surface, since heat will also be conducted through the superconductor. Also mitigating in favor of this approximation is that the thermal conductivity may be somewhat reduced near a free surface--the more so the longer the mean free path of the heat carrier. At these low temperatures, the phonons have a much longer mean free path than the electrons. If the normal region is a fluxoid, its field penetration into the neighboring superconducting region may also suppress $K_s$ for certain values of H and T. Yet, correction must be made for the approximations.

The analysis is directed to the case of an oscillating fluxoid which, as before, gives the largest power dissipation. A stationary normal region may be treated as mentioned in See. I.A. The solution to all cases may be put into the same general form so that the final results are quite general.

The average power loss/(cross-sectional area) down to any point x is:
$$P_a(x) = \tfrac{1}{2}\rho j_0^2 \int_0^x \exp(-2x/\lambda)dx$$
$$= \tfrac{1}{2}(\rho/2\lambda)H_p^2[1-\exp(-2x/\lambda)], \tag{1.13}$$
$$= \tfrac{1}{2}RH_p^2[1-\exp(-2x/\lambda)]$$

where
$$R \equiv \rho/2\lambda. \tag{1.14}$$

The heat flow equation is $P_a = -K_n'(dT/dx)$, or:
$$\tfrac{1}{2}RH_p^2[1-\exp(-2x/\lambda)] = -k_{1n}T(dT/dx), \tag{1.15}$$

The solution to Eq. (11. 5) is:

$$H_p = F^{-1} \left[ \frac{k_{1n}(T_a^2 - T_b^2)}{NRb} \right]^{1/2} , \qquad (1.16)$$

where $N \ll 1$, is put in as the correction factor for this case, and

$$F_1 \equiv \{1 - (\lambda/2b)[1 - \exp(-2b/\lambda)]\}^{1/2} \approx 1 \text{ for } b \gg \lambda. \qquad (1.17)$$

For $k_{1n}T_a^2 \gg k_{1n}T_b^2$,

$$H_p \approx F_1^{-1}[k_{1n}/NRb]^{1/2} T_a \left\{ 1 - \tfrac{1}{2}\left[k_{1n}T_b^2/k_1 T_a^2\right] - \tfrac{1}{8}[\ldots]^2 - \ldots \right\}. \qquad (1.18)$$

## II. MICROWAVE CAVITY Q

From Eqs. (1.12) and (1.18), we see that for both cases at large $T_a$, $T_a$ is approximately linear with $H_p$. If the superconductor is a microwave cavity and $T_a > \tfrac{1}{2}T_c$, then the cavity $Q \propto \exp(\varepsilon/2k_b T)$ where $\varepsilon$ is the energy gap and $k_B$ is the Boltzmann constant. If the cavity Q is dominated by the power loss around the cylinder (there may be more than one such normal region), $T \sim T_a \propto H_p$ under the conditions of (1.11) and (1.18),

$$\Rightarrow Q \propto \exp(D/H_p) \text{ for } H_p \text{ near } H_p', \qquad (2.1)$$

where, for the parallel case,

$$D \approx \frac{\varepsilon}{2k_B F} \left[ \frac{k_1}{NRa \, \ell n(b'/a)} \right]^{1/2} , \qquad (2.2)$$

and for the perpendicular case,

$$D \approx \frac{\varepsilon}{2k_B F_1} \left[ \frac{k_{1n}}{NRb} \right]^{1/2} . \qquad (2.3)$$

Expanding Eq. (2.1) to first order for $D\left[(1/H_p) - (1/H_p')\right] \ll 1$, gives

$$Q \approx \left[Q' - (Q'D/H_p')\right] + (Q'D/H_p), \qquad (2.4)$$

where $Q'$ is the cavity Q immediately prior to breakdown, and $H_p'$ is the magnetic breakdown field corresponding to $Q'$. When $T_a < \tfrac{1}{2}T_c$, then for $H_p$ near $H_p'$,

$$\Rightarrow Q \propto T \exp(\varepsilon/2k_B T) \propto H_p \exp(D/H_p) \qquad (2.5)$$

Expanding Eq. (2.5) to first order for $D\left[(1/H_p) - (1/H_p')\right] \ll 1$, yields

$$Q \approx (Q'D/H_p') + (Q'/H_p')\left[1 - (D/H_p')\right]H_p \tag{2.6}$$

Equation (2.6) gives a good representation of the data of Turneaure and Viet [5] as shown in Fig. 5.

$$D/H_p' \approx \frac{\varepsilon}{2k_B T_a} = AT_c/2T_a > 1, \tag{2.7}$$

where $\varepsilon = Ak_B T_c$ and $A \sim 4$.

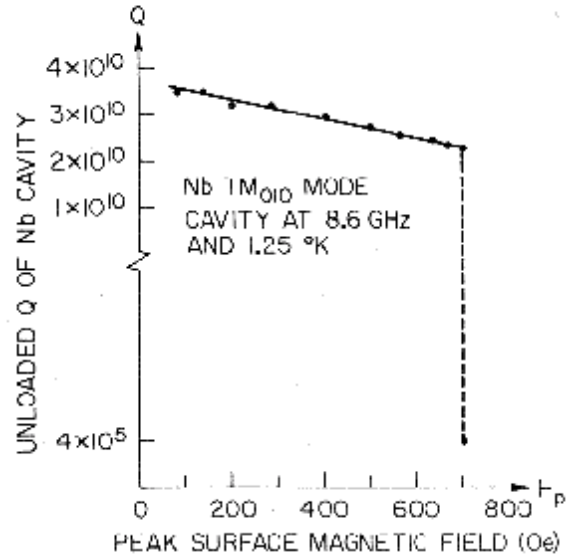

Fig.5. The unloaded Q of a Nb $TM_{010}$ mode cavity as a function of peak surface magnetic field, $H_p$. From Turneaure and Viet. [5]

Equations (2.4) and (2.6) were derived under the assumption that the normal region remains at a fixed equilibrium position. If the normal region is a fluxoid with components of flux perpendicular to the alternating current, it will oscillate about an equilibrium position. However, if it is weakly pinned, while oscillating, it will tend to migrate away from the regions of high current density due to current gradients in the superconductor until it becomes strongly pinned. If this happens, Q will first increase with increasing $H_p$, before decreasing. It should also be noted that there may well be a hysteresis-like effect when $H_p'$ is exceeded due to trapping of additional flux, and when $H_p$ is reduced, the original Q vs $H_p$ curve may not be obtained. When Q as a function of $H_p$ is measured in the temperature region corresponding to residual Q, it might appear that an equation like (2.5) would not be appropriate, as the residual Q is essentially independent of T. However, it should be noted that the temperature of the superconductor in the vicinity of the normal region may be significantly higher than the bath temperature and that of the rest of the cavity, giving Q its usual temperature dependence in this neighborhood. Prior to breakdown, conditions may be such that the region around the normal cylinder can only dominate the power loss when the rest of the cavity is operating near Q residual. The predictions made by Eqs. (2.4) and (2.6) are independent of any breakdown criterion, and of the nature of the normal region.

### III. MAGNETIC BREAKDOWN CRITERIA

When $H_p$ is increased to a point where the critical magnetic field is reached in the neighborhood, then the material surrounding the cylinder will go normal, leading to a sharp rise in the power dissipation and ultimately a run-away situation. This may be viewed as either a magnetic or thermal instability, as the two are linked together. In the case of a cavity, the Q will drop precipitously. The magnetic field in the neighborhood is $F\vec{H}_p \cos\omega t + \vec{H}_a$, where $\vec{H}_a$ is the magnetic field which exists at radius *a*, due to the contribution from all sources besides the current at the cylinder. (F = 1 for the perpendicular fluxoid.) $\vec{H}_a$ may be an applied dc field, and/or the field penetration from a fluxoid. The worst case is when $\vec{H}_a$ and $F\vec{H}_p \cos\omega t$ add together algebraically at peak value.

Type I:     $H_a + FH_p = H_c = H_o \left[1 - (T_a / T_c)^2 \right]$ (3.1)

For type II, breakdown will occur when the second critical field, $H_{c2}$, is exceeded in the neighborhood, causing it to go completely normal. Various relationships may be found

in the literature for $H_{c2}$ as a function of temperature. The following will suffice for our purposes:

Type II: $H_a + FH_p = H_{c2} = H_o \left[ \dfrac{1-(T_a/T_c)^2}{1+(T_a/T_c)^2} \right] \approx H_o \left[ 1 - 2(T_a/T_c)^2 \right]$ for $T_a \ll T_c$, (3.2)

or

$$H_a + FH_p \approx H_o \left[ 1 - B(T_a/T_c)^2 \right],\qquad(3.3)$$

where $B = 1$ for type I, and $B = 2$ for type II.

The simultaneous solution of Eqs. (1.9), (1.10), and (3.3) yields $H_p'$, $T_a$, and $T_f$ for the parallel region. If the substrate is not too thick and is a good thermal conductor, such as copper, $T_f \approx T_b$, and the result is the same as if there were no substrate. In either case, the solution for the magnetic breakdown field of the parallel normal region is:

$$H_p' = \dfrac{\left( \dfrac{-k_1 T_c^2}{2BH_o} + \left\{ \left( \dfrac{k_1 T_c^2}{2BH_o} \right)^2 - 2NRa\,\ell n\!\left(\dfrac{b'}{a}\right)\left[ \tfrac{1}{4}k_3\left(T_b^4 - h^4 T_c^4\right) + \tfrac{1}{2}k_1 T_c^2\left( g^2 - \dfrac{1}{B} + \dfrac{H_a}{BH_o} \right) - C_1 \right] \right\}^{1/2} \right)}{NFRa\,\ell n\!\left(\dfrac{b'}{a}\right)}$$

(3.4)

where $C_1 = c_2 T_c (g-h) + 2k_2 T_c^{-1/2}\left[ h^{-1/2} - g^{-1/2} \right]$.

Similarly, the simultaneous solution of Eqs. (1.16) and (3.3) gives for the perpendicular fluxoid:

$$H_p' = \dfrac{\left( \dfrac{-k_{1n} T_c^2}{2BH_o} + \left\{ \left( \dfrac{k_{1n} T_c^2}{2BH_o} \right)^2 - 2NRF_1^2 b\left[ \tfrac{1}{2}k_{1n}\left\{ T_b^2 - \dfrac{T_c^2}{B}\left(1 - \dfrac{H_a}{H_o}\right) \right\} \right] \right\}^{1/2} \right)}{NRF_1^2 b}$$ (3.5)

Equations (3.4) and (3.5) give the peak magnetic field at the surface of a superconductor which produces a critical power dissipation leading to a steep rise in the power loss whose origin is the heating of a normal region. A graphical interpretation of the solution for $H_p'$ is given in Fig. 6. Though the two equations are quite similar, they differ sufficiently that it should be possible to distinguish $H_p'$ experimentally between the two cases. Since the two cases also differ in total power dissipation, their effect on cavity Q should also be distinguishable.

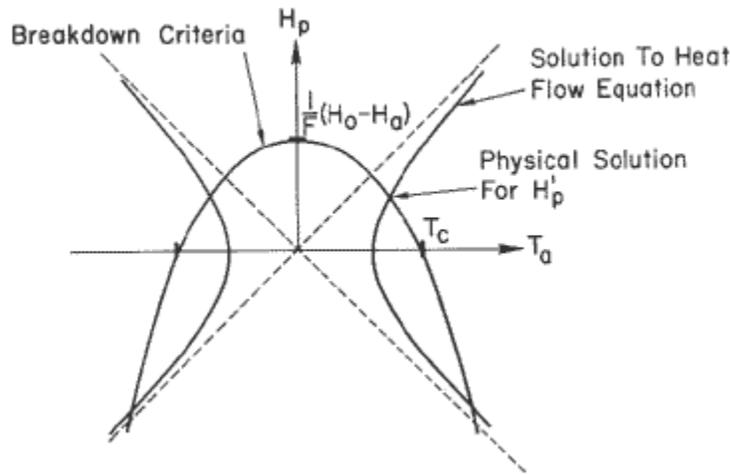

Fig.6. A graphical interpretation of the solution for the magnetic breakdown field, $H_p'$.

There are several competing mechanisms, as mentioned earlier, which can be responsible for magnetic breakdown. The mechanism analyzed in this paper may not be the dominant one in a given situation. If the normal region lies far below the conducting surface, resulting in a small value for F, Eq. (3.4) may even yield a value of $H_p'$ greater than the critical field. Equations (3.4) and (3.5) predict the breakdown field when the thermal lowering of the critical field due to power dissipation in a normal region is the dominant mechanism. The normal region was chosen parallel or perpendicular to the conducting surface to simplify the theoretical analysis. Aside from changes in geometrical factors, the basic relationships between the physical parameters should remain substantially the same for other geometries.

As mentioned briefly in Sec. II, there may be a hysteresis-like effect when $H_p$ is reduced following breakdown. This may be due to two factors. The regions which were driven normal during breakdown, will remain normal at much lower values of $H_p$ than were required to drive them normal, due to their enhanced power dissipation. In addition there may be trapping of additional flux in these regions.

## IV. EFFECTIVE RESISTIVITY

### A. Oscillating Fluxoid

Gittleman and Rosenblum [6, 7} have theoretically analyzed the case of a type II superconductor in the mixed state containing a rigid lattice of flux tubes or fluxoids. Bardeen and Stephen [8] have focussed their analysis on the unidirectional motion of a single flux tube in the array for a type II superconductor. What is essential here is the case of at least a single isolated oscillating fluxoid trapped in either a type I or type II superconductor due to an incomplete Meissner-Ochsenfeld effect as the superconductor is cooled below its transition temperature. The method will correspond more to that of Gittleman and Rosenblum, though the problem considered will be somewhat similar to that of Bardeen and Stephen. Kim, Hempstead, and Strnad [9] have done considerable work on d. c. flux flow in type II superconductors. The experimental work of Tholfsen and Meissner [10] on dc flux flow in type I superconducting films leads them to agree with the considerable experimental and theoretical evidence referenced in their paper that fluxoid formation and flow in type I films is similar to that in type II materials. However, they have a point of disagreement with Kim et al. [9] in which they claim a better fit to both their data and that of Kim et al. with a slightly different mathematical relationship. The relevance of their disagreement to this analysis will be pointed out shortly. In Maki's [11] theoretical treatment of unidirectional vortex motion in type I superconductors, he concludes that his "results have a close resemblance to those we have for vortex motion in the type II superconductor. We have the same expression for the electric current as Eq. (27) in the Abrikosov vortex state." His result is in terms of the diffusion constant of the condensed electron pair, and so does not lend itself readily to the analysis here. The analysis given is for the case of the parallel fluxoid. The analysis and the results would be essentially the same for the perpendicular fluxoid.

The following forces per unit length act on the isolated fluxoid:

Lorentz force = $\text{Re}\left[j_o' \exp(i\omega t)\phi\right]$, (4.1)

where the spatial variation of the superconducting transport current density is neglected, $j = j_o \exp[-(d+X)/\lambda]\exp(i\omega t) \approx j_o'\exp(i\omega t)$, $j_o' \equiv j_o \exp-(d/\lambda)$; $x = d + X$. $\phi$ is the total flux trapped in the fluxoid, $\phi = q\phi_o$ where $\phi_o = h/2e$ is the flux quantum and q is an integer $\geq 1$. $j$ is taken perpendicular to H, the magnetic field in the fluxoid, of permeability $\mu$.

$$H = \phi/\pi a^2 \mu.$$ (4.2)

For type II, $H_{c1} \leq H \leq H_{c2}$; and for type I, $H = H_c$.

Damping force = $\eta(dX/dt) = (\mu\phi H_o/\rho_n)(dX/dt)$, (4.3)

where $\eta$ is the flow viscosity empirically found by Kim, Hempstead, and Strnad. [9] $\rho_n$ is the normal-state resistivity, which may be taken to be the residual resistivity for T < 4° K.

$$\begin{aligned}H_o &\equiv H_c(0) \quad \text{for type I} \\ &\equiv H_{c2}(0) \quad \text{for type II}\end{aligned}$$ (4.4)

If we convert the Tholfsen and Meissner [10] data fitting result into a fluxoid flow viscosity, $\eta$, the result would be

$$\eta = \mu\phi[H_c(T)]^{3/2}/0.76 H^{1/2}\rho_n.$$ (4.5)

Since the Kim et al. equation for $\eta$ is generally accepted, it is the one that will be used here. However, in what follows, either expression, or any other, may be inserted where $\eta$ appears.

With no a priori knowledge of the pinning force, it will be assumed to vary linearly which should be a good approximation to most pinning potentials for small amplitude oscillation:

$$\text{Pinning force} = -pX. \tag{4.6}$$

The equation of motion of the single fluxoid is thus:
$$M(d^2X/dt^2) + \eta(dX/dt) + pX = \text{Re}\left[\phi j_o' \exp(i\omega t)\right]. \tag{4.7}$$

The solution to Eq. (4.7) is
$$X = \text{Re}\{X_o \exp[i(\omega t + \delta)]\}, \tag{4.8}$$

where
$$X_o = \phi j_o'\left[(\omega^2 M - p)^2 + \eta^2 \omega^2\right]^{-1/2} \tag{4.9}$$

and
$$\delta = \tan^{-1}\left[\omega / (\omega^2 M - p)\right]. \tag{4.10}$$

The average power dissipation per unit volume is
$$dP'/dV = \tfrac{1}{2} \text{Re}\left[\vec{j} \times \mu \vec{H} \cdot (d\vec{X}^*/dt)\right]$$
$$= \tfrac{1}{2}\left(\frac{\omega^2 \phi^2 H H_o \rho_n \mu^2}{\rho_n^2(\omega^2 M - p)^2 + \omega^2 \phi^2 H_o^2 \mu^2}\right), \tag{4.11}, (4.12), (4.13)$$
$$= \tfrac{1}{2} \rho j_o'$$

where the effective resistivity of the oscillating fluxoid is
$$\rho = \left\{\omega^2 \phi^2 H H_o \mu^2 / \left[\rho_n^2(\omega^2 M - p)^2 + \omega^2 \phi^2 H_o^2 \mu^2\right]\right\}\rho_n. \tag{4.14}$$

If the viscous damping force dominates, $\omega^2 \phi^2 H_o^2 \mu^2 \gg \rho_n^2 (\omega^2 M - p)^2$, then
$$\rho = (H/H_o)\rho_n. \tag{4.15}$$

This is the same result as obtained empirically by Kim et al. [9] and analytically by Gittleman and Rosenblum [6] for the mixed state of a type II superconductor containing a lattice of fluxoids. This is because when the viscous force dominates, their results should be independent of any pinning potential periodicity related to the flux tube lattice. In this case, for a type I superconductor, $\rho$ ranges from
$$\left[1 - (T_a/T_c)^2\right]\rho_n \leq \rho \leq \rho_n. \tag{4.16}$$

For a type II superconductor,

$$\frac{H_{c1}(0)\left[1-(T_a/T_c)^2\right]}{H_{c2}(0)}\rho_n \leq \rho \leq \rho_n. \tag{4.17}$$

If the viscous damping force is negligible, $\omega^2\phi^2 H_o^2\mu^2 \ll \rho_n^2(\omega^2 M - p)^2$, then

$$\rho = \left\{\omega^2\phi^2 HH_o\mu^2 / \left[(\omega^2 M - p)^2\right]\right\}/\rho_n. \tag{4.18}$$

Bardeen and Stephen [8] have derived the effective mass per unit length of a fluxoid to be:

$$M = 2\pi n m a^2 (H_o/H)^2 \sin^2\alpha, \tag{4.19}$$

where n is the electron density, m is the electron mass, and $\alpha$ is the Hall angle. Suhl [12] has also calculated the effective mass of a flux tube. A different expression for M was derived by Gittleman and Rosenblum [7] which is not just the property of an isolated fluxoid but is more related to the presence of a flux tube lattice and is also more appropriate for materials in the dirty limit.

If we substitute Eq. (4.19) and (4.2) in Eq. (4.14), we have, for the general case,

$$\rho = \left(\frac{\omega^2\pi^2 a^4 H^3 H_o\mu^4}{\rho_n^2\left[2\pi\omega^2 n m a^2 (H_o/H)^2 \sin^2\alpha - p\right]^2 + \mu^4\omega^2\pi^2 a^4 H^2 H_o^2}\right)\rho_n. \tag{4.20}$$

No one has measured both the thermal and electrical conductivity of a cavity or even of a representative material sample. Published joint measurements for a given sample are difficult to find, if not altogether absent. Fortunately, the Weidemann-Franz law can help us to obtain the resistivity from the thermal conductivity. We need the superconducting thermal conductivity to apply the parallel case. If the normal-state thermal conductivity, $K_n^{'}$ of the superconductor is also available,

$$\rho_n = (L/K_n')T = (L/k_{1n}T)T = L/k_{1n}, \tag{4.21}$$

where $L = 2.45 \times 10^{-8}\ W \cdot \Omega/°K^2$. If the slope, $k_{1n}$, of the normal conductivity is not available, one may use $k_{1n} \approx k_1$ to obtain $\rho_n$; and to apply the perpendicular case.

When the viscous force dominates, giving $\rho \propto \rho_n$, this theory can help to explain why exposure of the clean superconducting material to contaminating air (which may form oxides, etc.) can decrease Q and $H_p'$. If the fluxoid is at, or crosses, the conducting surface which has been contaminated, the resulting increase in $\rho_n$ can help to account for the deterioration of Q and $H_p'$.

## B. Stationary Normal Region

Now that we have considered an oscillating fluxoid, we still need to consider the case where the fluxoid is either so strongly pinned that it cannot oscillate, or there is no Lorentz force acting on it because j is parallel to H. Since the effects of the fluxoid's magnetic field will be neglected in this case, the analysis will be equivalent to the presence of any stationary normal region in the superconductor. [13] It will be shown that the effective resistivity in this case is much less than for an oscillating fluxoid, and that it may be the source of residual surface resistance.

Using the same notation as in Section IV. A, the electric field in a superconductor due to the time variation of the current is:

$$E = (m/e)[(1/ne)(dj/dt)] = (im\omega/ne^2)j = \rho_s j, \tag{5.1}$$

where the superconducting electrical resistivity is

$$\rho_s = i(m\omega/ne^2). \tag{5.2}$$

Although the superconducting current density, j, is given, the current density in the normal region, $j_n$, is not necessarily the same. In the case of the oscillating fluxoid, the expression (4.12) indicates that the oscillating fluxoid is equivalent to a stationary fluxoid of effective resistivity $\rho$ with the supercurrent flowing uniformly through both the fluxoid and the surrounding superconducting region. The current in a stationary conductor flows in such a manner as to minimize resistive losses for dc or low frequencies. However, at high frequencies, as in a GHz cavity, the stored electromagnetic energy is minimized and the current tends to flow more uniformly through the conducting surfaces. This would give a larger power dissipation; however, let us find out the least we can expect for the power loss. Even though j, flowing parallel to the normal region, may not be the same as $j_n$, E is continuous across the superconducting to normal boundary if there is no net surface charge density and the permittivity is the same in both regions. The tangential component of E is continuous in all cases.

From Maxwell's equations,

$$\nabla^2 E_n = i\omega\mu\sigma'_n E_n, \qquad (5.3)$$

where $\mu$ is the permeability of the normal region, and $\sigma'_n$ is its effective conductivity. If the field $E_n$ inside the normal cylindrical region were parallel to its axis, the solution to Eq. (5.3) would be

$$E_n = \frac{E J_o\left[\left(\frac{2}{i}\right)^{1/2} \frac{r}{\delta}\right]}{J_o\left[\left(\frac{2}{i}\right)^{1/2} \frac{a}{\delta}\right]}, \qquad (5.4)$$

where $J_o$ is the Bessel function of the first kind and zero order, $\delta$ is the anomalous skin depth, and r is the radial distance from the cylinder axis. For $a = \delta$, $E_n$ is reduced by only 6% at its lowest point on the axis. Therefore, to a good approximation for $a \leq \delta$, $E_n$ may be considered to be roughly constant in the normal region; and similarly for $j_n$. This conclusion would be valid at any cross section parallel to E for any orientation of the cylinder with respect to the electric field.

Hence the average power loss per unit volume in the normal region is

$$dP'/dV = \tfrac{1}{2}\rho'_n j_{no}^2 = (1/2\rho'_n)|E_o|^2, \qquad (5.5)$$

where $\rho'_n$ is the effective resistivity of the normal region if we use the actual current in this region.

$$\rho_n' = G\rho_n, \qquad (5.6)$$

where the factor G ≥1 takes into consideration that the electron mean free path, $\ell$, may be large compared with the dimensions of the normal region. G will be derived shortly.

It is more desirable to find the power loss in the normal region in terms of the supercurrent. Hence substituting Eq. (5.1) into Eq. (5.5),

$$dP'/dV = \tfrac{1}{2}\left(|\rho_s|^2/\rho_n'\right)j_{no}'^2 = \tfrac{1}{2}\rho j_o'^2, \qquad (5.7)$$

where,

$$\rho = |\rho_s|^2/\rho_n' = \left(m^2\omega^2/n^2 e^4 G\right)(1/\rho_n) \qquad (5.8)$$

is the effective resistivity of the stationary normal region if the supercurrent were to flow through it. It is interesting to note the similarity between Eq. (5.8) and Eq. (4.18) when $p \gg \omega^2 M$, i. e., when the pinning force is very high.

$$\rho_n = m/ne^2\tau = mv_F/ne^2\ell, \qquad (5.9)$$

where $\tau$ is the relaxation time between electron collisions, and $v_F$ is the Fermi velocity of the electrons. Combining Eq. (5.8) and (5.9),

$$\rho = m\omega^2\tau/ne^2 G, \text{ or } \rho/\rho_n = \omega^2\tau^2/G. \qquad (5.10)$$

Now to find G. If the normal region is a cylinder of radius a, the effective mean free path, $\bar{\ell}_e$ of the electrons is

$$1/\bar{\ell}_e \sim 1/\ell + 1/2a = (\ell + 2a)/2a\ell. \qquad (6.1)$$

Now $\rho_n' \propto 1/\bar{\ell}_e$ as can be seen from Eq. (5.9), hence

$$\rho_n' = (\ell/\bar{\ell}_e)\rho_n \sim [1 + (\ell/2a)]\rho_n = G\rho_n \qquad (6.2)$$

$$G \sim [1 + (\ell/2a)]. \qquad (6.3)$$

We can get a similar result more formally as follows: Since $\rho_n' \propto 1/\bar{\ell}_e$, we want to find $\bar{\ell}_e$. If the electrons are scattered isotropically from the center line of the cylinder and $\ell > a$,

$$\ell_e = a/\cos\theta \quad 0 \leq \theta \leq \theta_o \equiv \cos^{-1}(a/\ell)$$
$$= \ell \quad \theta_o \leq \theta \leq \theta_o + \alpha, \ \alpha \equiv 2\sin^{-1}(a/\ell)$$
(6.4)

where $\theta$ is the angle of the scattered electrons' trajectory measured w.r.t. a line perpendicular to the cylinder's axis. Averaging over the solid angle $\Omega$,

$$\bar{\ell}_e = \int \ell_e d\Omega / \int d\Omega$$
$$= \frac{1}{4\pi}\left[2\int_0^{\theta_o}\left(\frac{a}{\cos\theta}\right)2\pi\cos\theta d\theta + 2\int_0^{\theta_o+(1/2)\alpha}2\pi\ell\cos\theta d\theta\right];$$
$$\approx a[(\pi/2)-(a/2\ell)]$$
(6.5)

$$G = \ell/\bar{\ell}_e$$
$$= \frac{\ell}{a[(\pi/2)-(a/2\ell)]}.$$
(6.6)

If $\ell \gg 2a$, $G = 2\ell/\pi a$ and the result is almost the same as given by Eq. (6.3). In this limit,

$$\rho = m\omega^2\ell/ne^2v_F G = m\omega^2\pi a/2n^2e^2v_F.$$
(6.7)

There will be an additional power loss from the normal electrons which leave the normal region and enter the surrounding superconducting region. However, since the power loss in the normal regfoin is relatively small, compared with an oscillating fluxoid, the power loss in the surrounding region should also be relatively small. Mitigating against this additional loss is that essentially only those electrons in the normal region with energies above the energy gap or energies corresponding to unfilled states below the gap can contribute to this loss. This restriction was neglected in calculating G, and would tend to make G smaller. Also neglected were induced eddy currents.

Insofar as the approximations made in deriving Eq. (6.7) are valid, the effective resistivity of the stationary normal region is independent of $\ell$, as long as $\ell \gg 2a$, and hence will be independent of purity and lattice defects. Since $\rho \propto a$, and the other parameters are easily obtained, this may be a way to determine *a*. This might be possible, using a superconducting cavity at frequencies below the pinning frequency. The apparatus would have to be quite sensitive as the power loss in this case will be considerably lower than for an oscillating fluxoid.

The circular cylindrical shape for the normal region was chosen as a first approximation. Due to the anisotropy of the thermal conduction and the gradient in power dissipation in the x-direction, there will be a temperature gradient across the fluxoid. Thus in the parallel case, it will deviate from a circular cross section. The top, being at the highest temperature, will have the largest radius of curvature and the bottom will have the smallest. The cross section will be somewhat "pear-shaped" as depicted by the fluxoid on the right side of Fig. 1. In the perpendicular case, the fluxoid will assume a truncated cone shape, as shown on the right side of Fig. 4. Nevertheless, the radius *a* determined from Eq. (6.7) should be a good approximation to the size of the fluxoid, or whatever the normal region may be.

An interesting observation can be made about Eq. (6.7) or any of its forms, such as Eq. (5.8) and (5.10). Similarly, for Eq. (4.18) when $p \gg \omega^2 M$. For very high Q superconducting cavities, the residual surface resistance, $R_{res}$ is independent of temperature and field level, is $\propto \omega^2$, and is present even for extremely small stored energy. [5, 14 - 16] As we have seen earlier, $R \propto \rho / \lambda$, hence $R_{res} \propto \rho / \lambda$. $\lambda$ is frequency independent, and essentially temperature independent for $T < \frac{1}{2} T_c$. Therefore, a stationary normal region would give $R_{res} \propto \omega^2$ and independent of T and power level. This fits the experimental findings, and should be looked into further. Induced eddy current power loss will also have an effective surface resistance with similar dependencies.

Residual power loss results from the contribution from all sources, whereas critical power loss is likely due to one region which first goes unstable. A cavity may have its residual Q deteriorate by orders of magnitude, whereas its breakdown field may only decrease by a small factor. Similarly, it is possible that a single fluxoid which may cause breakdown at high field levels, may contribute less at low levels than the combined loss from all other sources. Nevertheless, the speculation that the same source may be responsible for both residual and critical power loss is worthy of consideration. If a fluxoid went from an oscillating mode to a very small amplitude mode as $T \to 0$ and low field, then it could be the cause of both losses. The possibility that critical power loss is initiated by a nonmagnetic, but quite thermally isolated normal region, also deserves attention as this could also account for both losses.

## V. CONCLUSIONS

We have seen that a simplified but otherwise rather general model of a very small region of normal power dissipation can explain why the magnetic breakdown field, $H_p'$, can be significantly lower than the critical field. As shown earlier by the author [17], the theory predicts a reasonable range of breakdown fields, depending on the grade of niobium, caused by a single flux quantum trapped in a superconducting cavity. This range is in agreement with the measured magnetic breakdown fields of Turneaure and Weissman, [16] and with the subsequent measurements by Turneaure and Viet [5], which extended the breakdown fields primarily by improved vacuum heat treatment of the Nb cavities. The theoretical predictions of this paper are consistent with, and serve to explain, their findings as well as the fact that the thermal conductivity improves with vacuum heat treatment. Additionally, this theory predicts that for $H_p$ near $H_p'$, the cavity $Q \propto \exp(D/H_p)$ for $T_a > \frac{1}{2}T_c$, and for $T_a < \frac{1}{2}T_c$, $Q \propto H_p \exp(D/H_p)$. This is a good representation of the data of Turneaure and Viet. [5] As far as the author was able to determine, there is no other theory which predicts these experimental results.

From this model of the thermal nature of $H_p'$, not only can one explain why $H_p' < H_c$ ($H_{c1}$ for type II) and predict values of $H_p'$ in terms of material parameters, but also a reason for lower values of $H_p'$ for thin films than for the bulk material can also be given. In addition to effects of impurities and grain size, strain can also decrease thermal conductivity in the superconducting state. [4] Due to a lattice mismatch between film and substrate, strain can be induced in the film. As the theoretical analysis indicates, a reduction in thermal conductivity leads to a reduction in magnetic breakdown field. (There may also be other reasons for the low breakdown fields of thin films related to impurities, etc. , and the inability to give them a high temperature vacuum heat-treatment due to the low melting point of the substrate. )

As was shown, the effective resistivity, $\rho$, of a single isolated oscillating fluxoid has no frequency dependence when the viscous force dominates, is - $\rho_n$ and is much larger than $\rho$ for a stationary normal region. When the viscous force is negligible and $p >> \omega^2 M$, $\rho \propto \omega^2 / \rho_n$ which is the same as the stationary normal region. This dependency predicts a residual surface resistance $\propto \omega^2$ and independent of temperature and field level in accord with experimental observation.

**ACKNOWLEDGMENTS**

I wish to express my appreciation for helpful comments and suggestions by Perry Wilson, Thomas Di Stefano, Edward Garwin, Matthew Allen, and Bruce Rosenblum. Thanks are also due to Vi Smoyer for her excellence in typing the manuscript.   This work was supported by the U. S. Atomic Energy Commission.